\documentclass[11pt,preprint]{aastex}

\newcommand{\etal}{{\it et~al.}}

\usepackage{rotating}

\begin{document}

\title{Volatility of Sodium in Carbonaceous Chondrites at Temperatures
  Consistent with Low-Perihelia Asteroids}

\author{Joseph R. Masiero\altaffilmark{1}, Bj\"{o}rn J. R. Davidsson\altaffilmark{2},  Yang Liu\altaffilmark{2}, Kelsey Moore\altaffilmark{2}, Michael Tuite\altaffilmark{2}}

\altaffiltext{1}{Caltech/IPAC, 1200 E California Blvd, MC 100-22, Pasadena, CA 91125, USA, {\it jmasiero@ipac.caltech.edu}}
\altaffiltext{2}{Jet Propulsion Laboratory/California Institute of Technology, 4800 Oak Grove Dr., MS 183-301, Pasadena, CA 91109, USA}

\begin{abstract}

Solar system bodies with surface and sub-surface volatiles will show
observational evidence of activity when they reach a temperature where
those volatiles change from solid to gas and are released.  This is
most frequently seen in comets, where activity is driven by the
sublimation of water, carbon dioxide, or carbon monoxide ices.
However, some bodies (notably the asteroid (3200) Phaethon) show
initiation of activity at very small heliocentric distances, long
after they have reached the sublimation temperatures of these ices.
We investigate whether the sodium present in the mineral matrix could
act as the volatile element responsible for this activity.  We conduct
theoretical modeling which indicates that sodium has the potential to
sublimate in the conditions that Phaethon experiences, depending on
the mineral phase it is held in.  To test this, we then exposed
samples of the carbonaceous chondrite Allende to varying heating
events similar to what would be experienced by low perihelion
asteroids.  We measured the change in sodium present in each sample,
and find that the highest temperature samples show a significant loss
of sodium from specific mineral phases over a single heating event,
comparable to a day on the surface of Phaethon.  Under specific
thermal histories possible for Phaethon, this outgassing could be
sufficient to explain this object's observed activity. This effect
would also be expected to be observed for other low-perihelia
asteroids as well, and may act as a critical step in the process of
disrupting small low-albedo asteroids.

\end{abstract}

\section{Introduction}

The small bodies of the inner Solar system are generally grouped into
two phenomenological classes: comets and asteroids.  The main
difference distinguishing these two populations is whether or not the
body is observed to show activity in the form of emission of gas and
dust.  This divide is usually traced to whether the object contains
frozen sub-surface volatile materials such as water, carbon dioxide,
or carbon monoxide, and therefore is used as a proxy for where in the
Solar system the object formed with respect to the protoplanetary ice
line of each molecule.

Recently, this clear distinction has been called into question
by a newly recognized population of objects.  These objects, on
asteroid-like orbits, have been seen to show comet-like activity
\citep{jewittAIV}.  Referred to by various publications as Active
Asteroids, Main Belt Comets, or Active Main Belt Objects, this
population can be divided into: objects showing activity from impact
events, objects with activity consistent with disruption after
rotational spin up, and objects with activity consistent with the
presence of subsurface water ice.

However, there are a few objects that fail to fit into any of these
broad categories.  The most notable example is the asteroid (3200)
Phaethon, a near-Earth object (NEO) which has been shown to be active
at perihelion \citep{jewitt10,li13} and is orbitally associated with
the Geminid meteor stream \citep{gustafson89}.  The perihelion of
Phaethon is $q=0.14~$AU and its aphelion is $Q=2.4~$AU, making the
surface temperatures on the day-lit side well above the sublimation
temperature of water ice for the majority of its orbit.  Any ice
thermally coupled to the surface would be expected to have been
removed quickly after Phaethon's dynamical transfer into its current
orbit, and not survive to the present day.  \citet{maclennan20} have
shown that over the course of its orbital evolution Phaethon has
undergone numerous cycles through high peak temperatures like today,
which would have baked off any water near the surface.  This means
water ice cannot be driving the observed activity, which is seen only
in a short temporal period around Phaethon's perihelion. This short
burst of activity would not be the behavior expected if water ice was
driving the activity.  While water could be bound up in hydrated
silicates, experiments by \citet{garenne14} have shown that these
minerals in meteorites begin losing mass below $873$ K, well
below Phaethon's perihelion temperature.  Recent spectral observations
by \citet{takir20} have indicated an absence of hydrated minerals on
the surface, further supporting the conclusion that they are not the
source of activity.

Previous investigations have looked into the effects of the extreme
temperatures that Phaethon experiences at perihelion
\citep{delbo14,ryabova18,yu19,maclennan20}.  These studies have found
that fracturing, driven by thermal cycling, can efficiently fragment
boulders with Phaethon's expected surface composition.  This would act
as a source of dust that could eventually be ejected from the surface
to create the observed brightening and the associated meteor stream.
However, thermal fracturing alone is unlikely to produce sufficient
force to let this dust reach escape velocity and be emitted from
Phaethon.

We propose that the driver of the activity on Phaethon at perihelion
is the volatilization of sodium bounded in minerals such as sodalite
and nepheline distributed in the matrix.  Other than water ice, which
was lost early during Phaethon's transition to a near-Earth orbit,
sodium is one of the most volatile elements that is present at a
significant quantity in asteroidal materials
\citep{notsu78,asplund09,sossi18}.  Mercury, with a perihelion of
$q=0.31~$AU, is known to have an exosphere populated by vaporized
sodium \citep{cassidy15} which means that sodium volatilization does
occur in the near-Sun region.  It is important to note that
\citet{cassidy15} show that the temperature profile of Mercury's
exosphere implies that the primary mechanism of volatilization is
photon-stimulated desorption instead of thermalization.  They
attribute the lack of thermalized sodium to space weathering, which we
would expect to be less significant for Phaethon given that it is a
dynamically much younger object.

Additionally, the Geminid meteors have been observed to be depleted in
sodium compared to other meteor streams \citep{kasuga05,abe20}.
\citet{kasuga06} investigated the sodium content of meteor showers,
finding that while the current orbit of Phaethon does not reach the
condensation temperatures for common sodium minerals such as feldspar,
small particles will experience higher temperatures and may be
devolatilized that way.  They suggest that the sub-millimeter scale
Geminid meteors were depleted in sodium after being ejected from the
parent body.  However, we propose here the converse: that it is the
devolatilization of sodium-bearing minerals on Phaethon that promoted
ejection of the particles that became the Geminids, and thus they were
sodium-depleted during their formation.

To investigate this hypothesis, we conducted thermophysical modeling
of Phaethon as it orbits the sun to determine the feasibility of this
mechanism.  We then performed heating experiments on meteorite samples
as an analog for Phaethon to search for sodium loss.  Sodium content
was determined for a range of peak heating temperatures on individual
minerals within each sample, which were analyzed post-heating for
changes to their chemistry.  This work is intended to be an initial
look into the possibility of sodium volatilization causing activity on
low-perihelion bodies, and future work will constrain this effect in
further detail.

\section{Thermophysical modeling of Phaethon}

To test the feasibility of our hypothesis that sodium sublimation is
driving the activity of Phaethon, we performed thermophysical modeling
of that process. We consider Phaethon as a spherical, rotating, and
orbiting body consisting of a porous mixture of silicate and sodium
grains. Constant model parameters are provided in
Table~\ref{tab.model1}. We apply the thermophysical model
\texttt{NIMBUS} (Numerical Icy Minor Body evolUtion Simulator)
developed by \citet{davidssonNIMBUS} \citep[see also][]{davidsson21}.
\texttt{NIMBUS} considers a number of latitudinal slabs (here 18,
evenly distributed in cosine), each split into a number of radial
cells, stretching from the body center to the surface (here 123 cells,
with a thickness of $200\,\mathrm{m}$ at the core, diminishing with
geometric progression to $0.003\,\mathrm{m}$ at the surface). For this
grid, \texttt{NIMBUS} solves a coupled system of differential
equations that govern the radial and latitudinal transport of heat and
mass throughout the body. The upper boundary condition of the energy
conservation equation balances absorbed Solar radiation (calculated as
a function of orbital position, latitude, and rotational phase) with
thermal reradiation into space, and heat conducted to or from the
surface. At depth, the energy conservation equation accounts for solid
state conduction, radiative transfer of energy in pores, energy
consumption of sublimating solid sodium, energy transport (advection)
of diffusing vaporized sodium, and energy release when vaporized
sodium recondenses into a solid.  We note that here we use diffusion
to refer to gas flowing through an empty channel (e.g. pore space
within the body), as opposed to atomic transport within solids which
we will refer to as `Fickian diffusion'.  The mass conservation
equation ensures that vapor diffuses according to local temperature
and pressure gradients, it tracks the withdrawal of the sodium
sublimation front below the surface, and regulates the release of
vaporized sodium into space.

\begin{table}[hb]
\begin{center}
\footnotesize
\begin{tabular}{||l|l|r|r|l||}
\hline
\hline
Symbol & Description & Value & Unit & Reference\\
\hline
$a$ & Semi--major axis & 1.27136 & $\mathrm{AU}$ & MPC\\
$e$ & Eccentricity & 0.8898495 & -- & MPC\\
$i$ & Inclination & 22.25964 & $^{\circ}$ & MPC\\
$\omega$ & Argument of perihelion & $322.18686$ & $^{\circ}$ & MPC\\
$\Omega$ & Longitude of the ascending node & $265.21748$ & $^{\circ}$ & MPC\\
$\mathcal{T}$ & Perihelion date & 2459191.0935865 & $\mathrm{JD}$ & MPC\\
\hline
\hline
$P$ & Rotation period & 3.603958 & $\mathrm{h}$ & \citet{hanus16}\\
$\lambda$ & Spin axis ecliptic longitude & 319 & $^{\circ}$ & \citet{hanus16}\\
$\beta$ & Spin axis ecliptic latitude & $-39$ & $^{\circ}$ & \citet{hanus16}\\
\hline
\hline
$D$ & Diameter &  4.6 & $\mathrm{km}$ & \citet{masiero19}\\
$A$ & Bond albedo & 0.04 & -- & \citet{golubeva20}\\
$\varepsilon$ & Emissivity & 0.9 & -- &\\
$\rho_{\rm sil}$ & Silicate grain density & 3000 & $\mathrm{kg\,m^{-3}}$ & \citet{macke11}\\
$\rho_{\rm Na}$ & Sodium grain density & 968 & $\mathrm{kg\,m^{-3}}$ & \\
$\psi$ & Porosity & 0.5 & -- & \\
$m_{\rm Na}$ & Sodium abundance & 0.5 & $\mathrm{wt\%}$ & \citet{lodders03}\\
$r_{\rm g}$ & Grain radius & 100 & $\mathrm{\mu m}$ & \\
$L_{\rm t}$ & Tube length & 1 & $\mathrm{mm}$ & \\
$r_{\rm t}$ & Tube radius & 100 & $\mathrm{\mu m}$ & \\
$\xi$ & Tortuosity & 1 & -- & \\
$T_0$ & Initial temperature & 260 & $\mathrm{K}$ & \\
\hline 
\hline
\end{tabular}
\caption{Orbital, rotational, and physical parameters applied in the thermophysical model of Phaethon. MPC is the Minor Planet Center (\it https://minorplanetcenter.net).}
\label{tab.model1}
\end{center}
\end{table}

A number of temperature--dependent and material--specific functions
are used by \texttt{NIMBUS} to regulate the behavior of the physical
processes under consideration. For the heat capacity $c=c(T)$ of both
rock and sodium it uses forsterite ($\mathrm{Mg_2SiO_4}$) as an
analog, as measured by \citet{robie82}.  The heat conductivity of
compacted material $\kappa=\kappa(T)$ is taken as that of the H5
ordinary chondrite Wellman measured by \citet{yomogida83}.  Because it
has been suggested that CK4 carbonaceous chondrites are Phaethon
analogs \citep{clark10} we apply the whole--rock density $\rho_{\rm
  sil}=3000\,\mathrm{kg\,m^{-3}}$ as measured for such meteorites
\citep{macke11}.  We correct the heat conductivity for porosity $\psi$
by applying the generation $n=2$ Hertz factor $h=h(\psi)$ function of
\citet{shoshany02}. This allow us to calculate an instantaneous
thermal inertia,
\begin{equation} \label{eq:01}
\Gamma=\sqrt{c(T)\rho(1-\psi)\kappa(T)h(\psi)}.
\end{equation}
Trial runs indicated that $\psi=0.5$ would yield
$670\stackrel{<}{_{\sim}}\Gamma\stackrel{<}{_{\sim}}1000\,$  J m$^{-2}$
s$^{-0.5}$ K$^{-1}$ for the surface temperature range experienced by
Phaethon. This is similar to the $\Gamma=600\pm200\,$J m$^{-2}$
s$^{-0.5}$ K$^{-1}$ range measured for Phaethon by \citet{hanus16} and
the $\Gamma=880^{+580}_{-330}\,$J m$^{-2}$ s$^{-0.5}$ K$^{-1}$ range
measured for Phaethon by \citet{masiero19}, which motivates our choice
of porosity.

To describe sodium sublimation we applied the latent heat and
Hertz--Knudsen formula parameters tabulated by \citet{huebner70}, that
allowed us to define the corresponding saturation partial
pressure ($\mathrm{[Pa]}$) of sodium vapor above metallic
  sodium:

  \begin{equation}
p_{\rm sat}(T)=10^{1.7372\cdot (\mathcal{L}-10)}   10^{\alpha-\beta L/T}\sqrt{\frac{2\cdot 10^{-3}\pi k_{\rm B}}{N_{\rm A}}}
\end{equation}

\noindent where $\alpha=29.770$ and $\beta=0.21860$ are constants,
$L=2.46\cdot 10^4\,\mathrm{cal\,mole^{-1}}$ is the latent heat,
$\mathcal{L}=L/R_0T_{\rm s}=10.8$ (where $R_0$ is the universal gas
constant and $T_{\rm s}$ is the sodium boiling temperature), $k_{\rm
  B}$ is the Boltzmann constant and $N_{\rm A}$ is the Avogadro
constant \citep{huebner70}.  Whenever the local partial pressure of
sodium vapor is below $p_{\rm sat}(T)$ it will trigger sublimation if
solid sodium is present. Conversely, sodium gas will re-condense where
the local partial pressure exceeds $p_{\rm sat}(T)$.  We consider pure
sodium here as the limiting case for what is occurring on Phaethon; in
reality any sodium would initially be contained in a silicate mineral
phase.  The volume mass sublimation and condensation rates are
calculated using standard expressions
\citep[e.g.][]{mekler90,prialnik92,tancredi94}. Gas diffusion fluxes
are calculated using the Clausing formula, evaluated as in
\citet{davidsson02}.  These expressions require a number of
geometrical parameters ($r_{\rm g}$, $L_{\rm t}$, $r_{\rm t}$, $\xi$),
for which we apply educated guesses (Table~\ref{tab.model1}).
At the modeled sublimation rate, the solution is not strongly
  sensitive to the diffusivity value that result from these
  geometrical parameters, as discussed by \citet{davidssonNIMBUS}. If
  the diffusivity changes, the temperature and pressure profiles
  adjust in such a way that the same net mass loss rate takes place
  and the same latent energy is being consumed.

The simulation was initiated at aphelion ($Q=2.40\,\mathrm{AU}$),
assuming an initial temperature $T_0=260\,\mathrm{K}$, and terminated
after the following perihelion ($q=0.14\,\mathrm{AU}$). This provided
sufficient time for the near--surface region to lose memory of the
arbitrary initial conditions. We also assumed that the (chondritic)
sodium abundance $m_{\rm Na}$ initially was constant throughout the
body.

\begin{figure}
\centering
\begin{tabular}{cc}
\scalebox{0.4}{\includegraphics{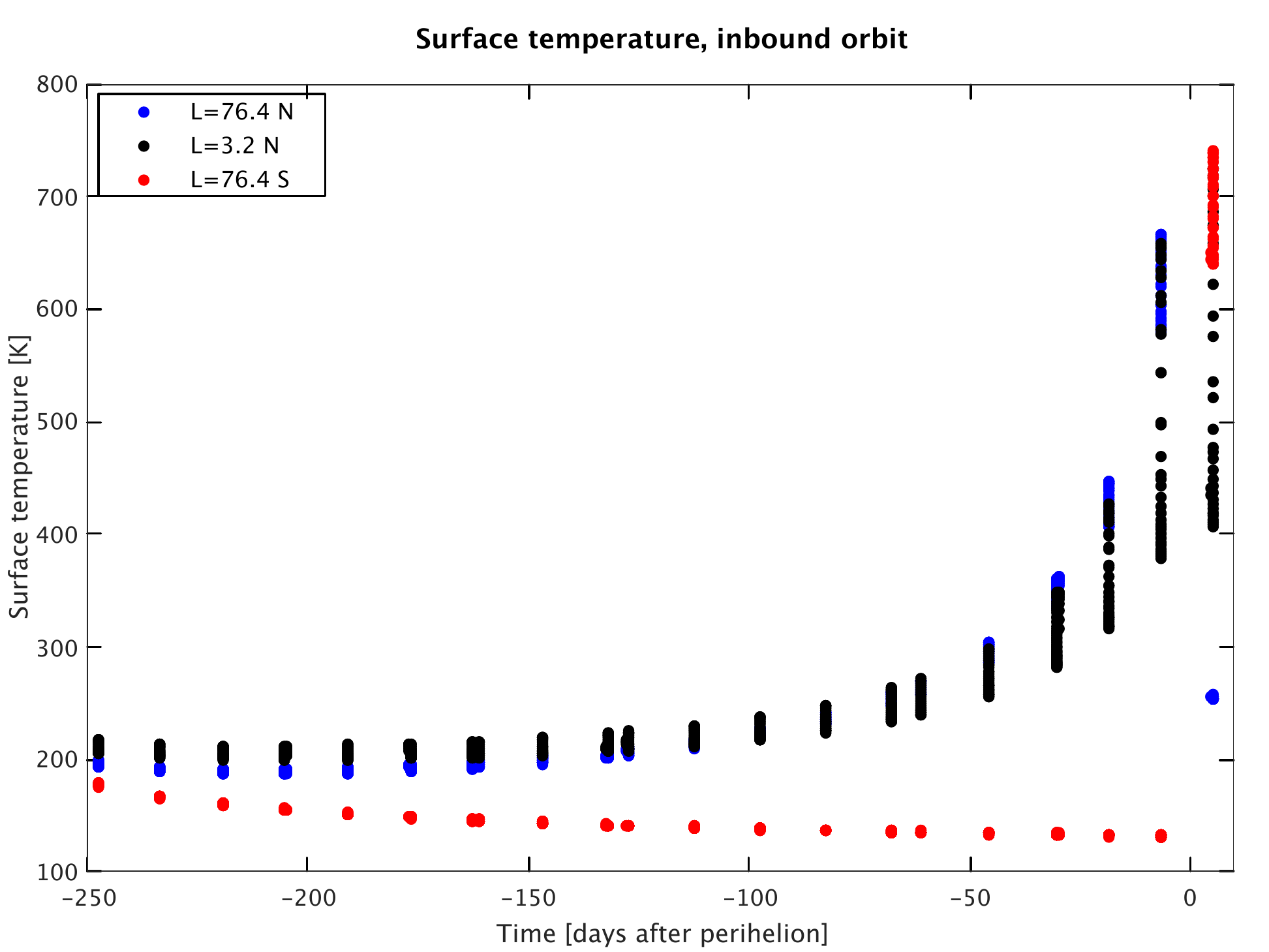}} & \scalebox{0.4}{\includegraphics{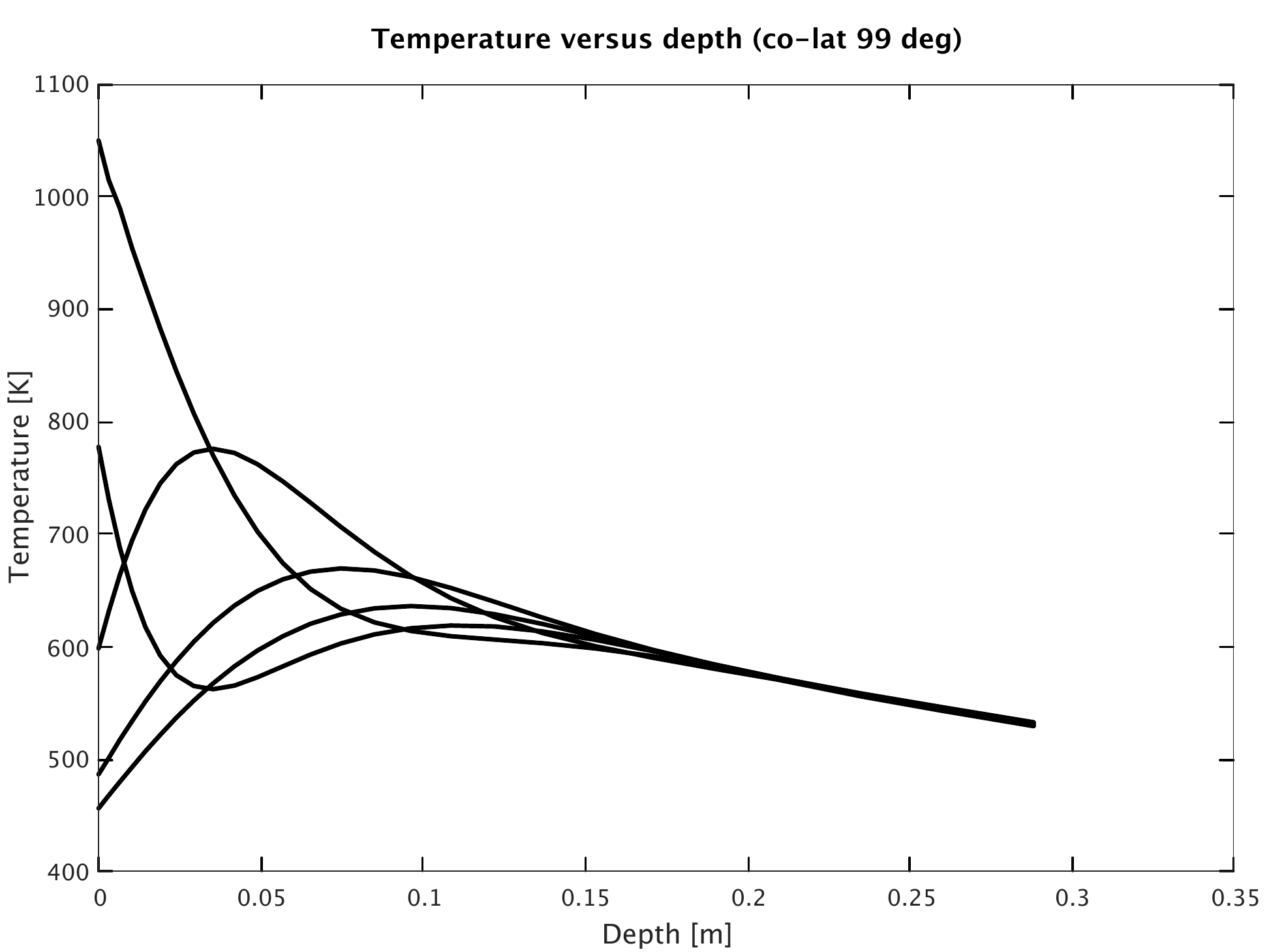}}\\
\end{tabular}
     \caption{Left: surface temperature at three specific latitudes as
       a function of time from aphelion to perihelion. The vertical
       extension of data--point clusters illustrate the range of
       day--night temperature variations. One rotation period every
       second week is plotted for clarity. Right: temperature as
       a function of depth for a handful of rotational phases, valid at
       perihelion for latitude $9^{\circ}\,\mathrm{S}$. Note that this
       particular revolution is not plotted in the left figure.}
     \label{fig.model1}
\end{figure}

Figure~\ref{fig.model1} shows that the surface temperature remains
below $\sim 300\,\mathrm{K}$ at all latitudes, except during a four
month period around perihelion.  Within four weeks of perihelion, the
peak temperature exceeds $\sim 700\,\mathrm{K}$ and briefly reaches a
maximum of $1050\,\mathrm{K}$ near the equator in the upper
  few centimeters of regolith as shown on the right in Fig
  \ref{fig.model1}.  These results are comparable to a recently
published thermophysical analysis of Phaethon by \citet{maclennan20}
and \citet{abe20}.  Interestingly, the spin axis is oriented such that
the northern hemisphere is illuminated up to perihelion. The equinox
occurs close to this point which keeps the south pole dark and cold
until it suddenly receives substantial illumination. The drastic
temperature jump of the south pole from nearly $100\,\mathrm{K}$ to
above $700\,\mathrm{K}$ over the course of a week could potentially
explain the surge of activity seen near perihelion
\citep{jewitt10,li13}.

\begin{figure}
\centering
\begin{tabular}{cc}
\scalebox{0.3}{\includegraphics{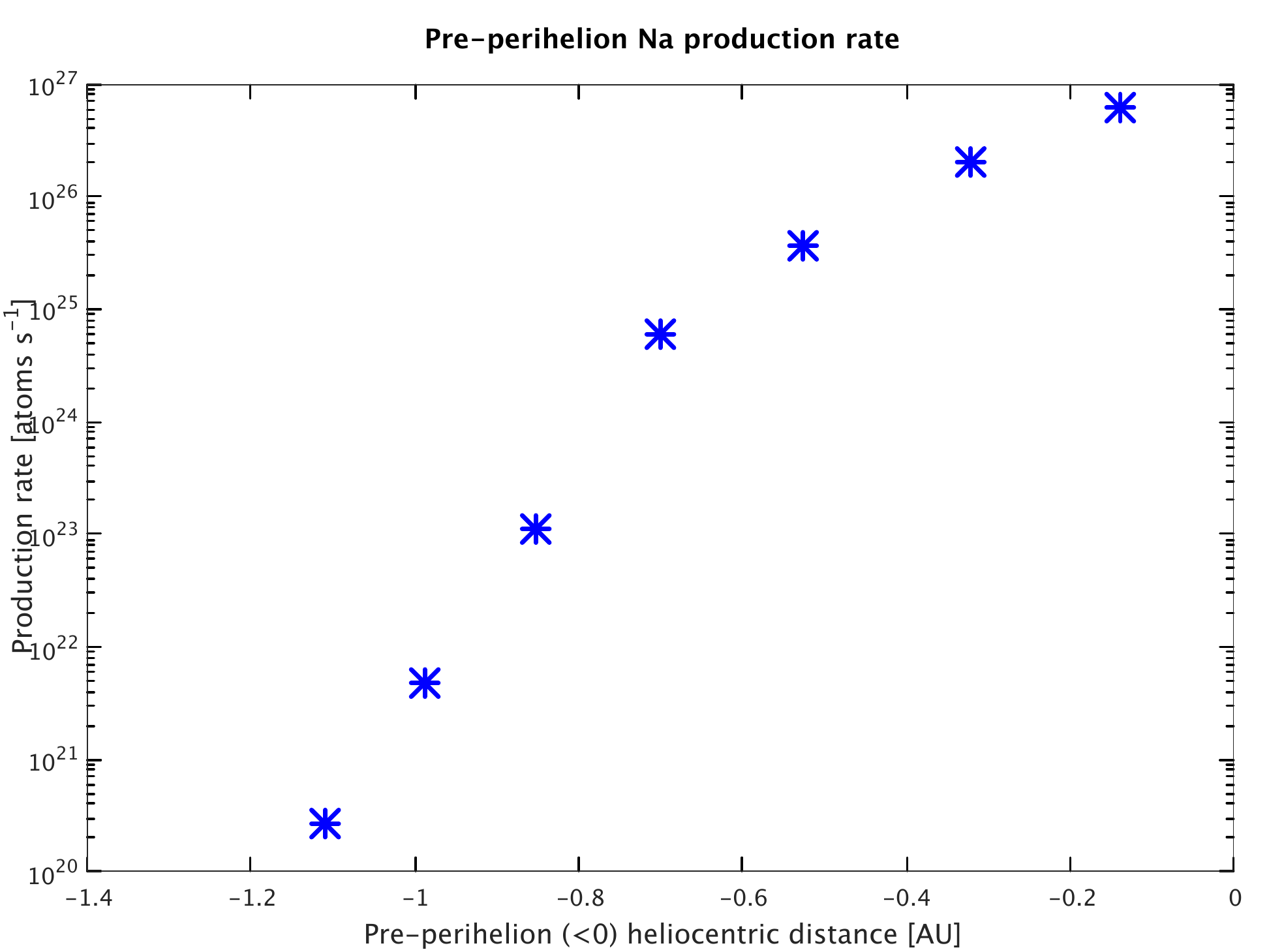}} & \scalebox{0.3}{\includegraphics{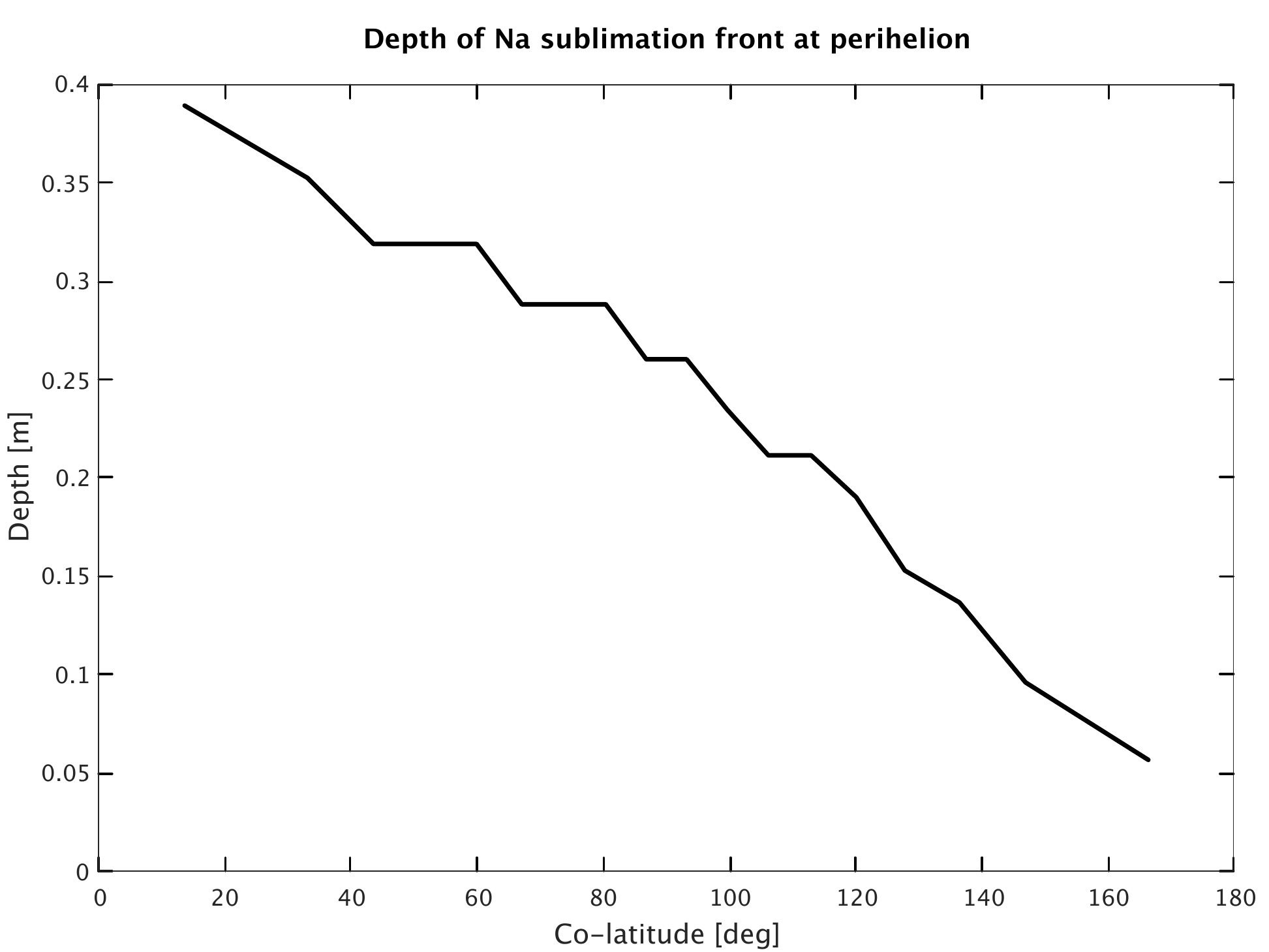}}\\
\scalebox{0.3}{\includegraphics{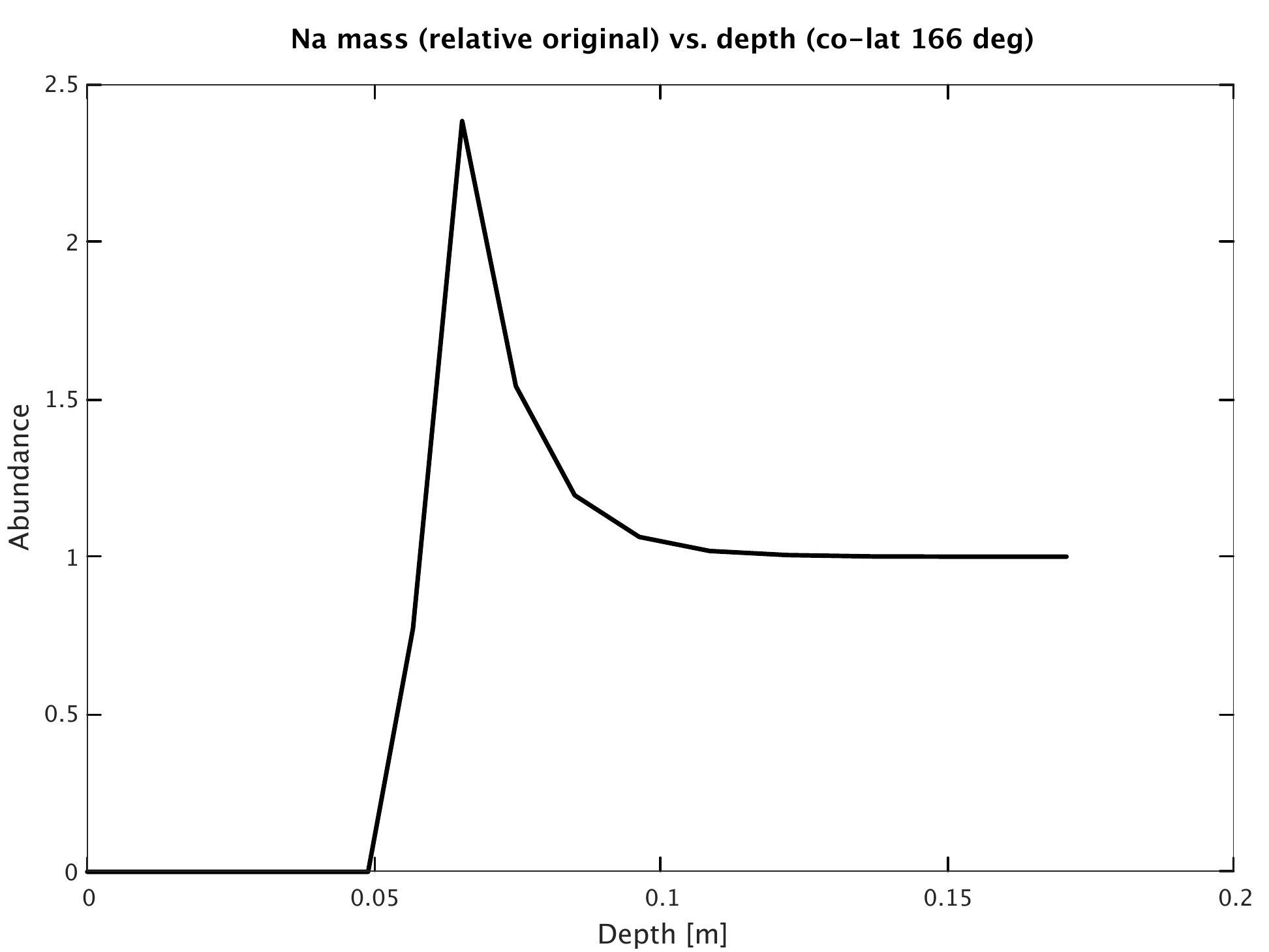}} & \scalebox{0.3}{\includegraphics{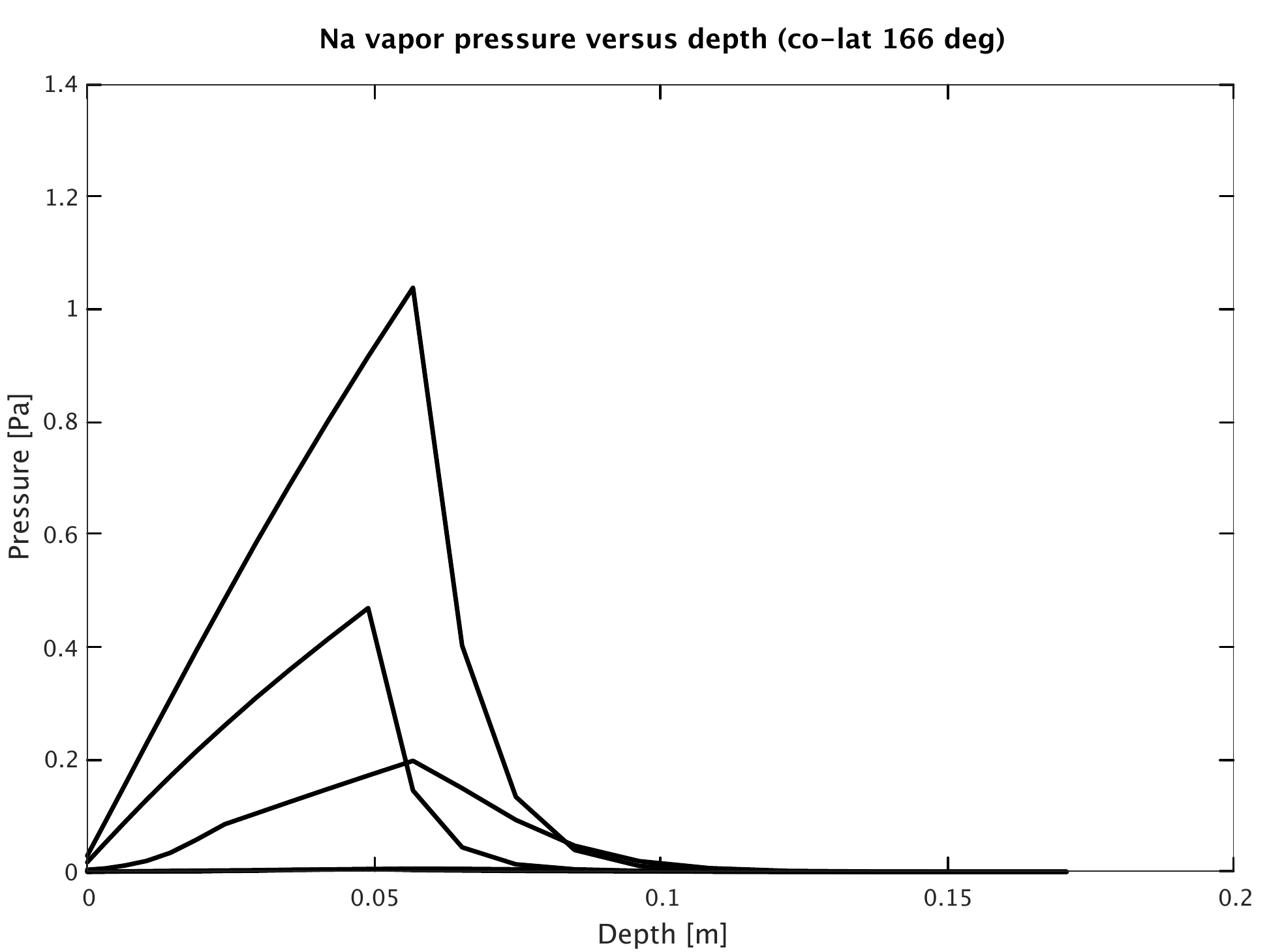}}\\
\scalebox{0.3}{\includegraphics{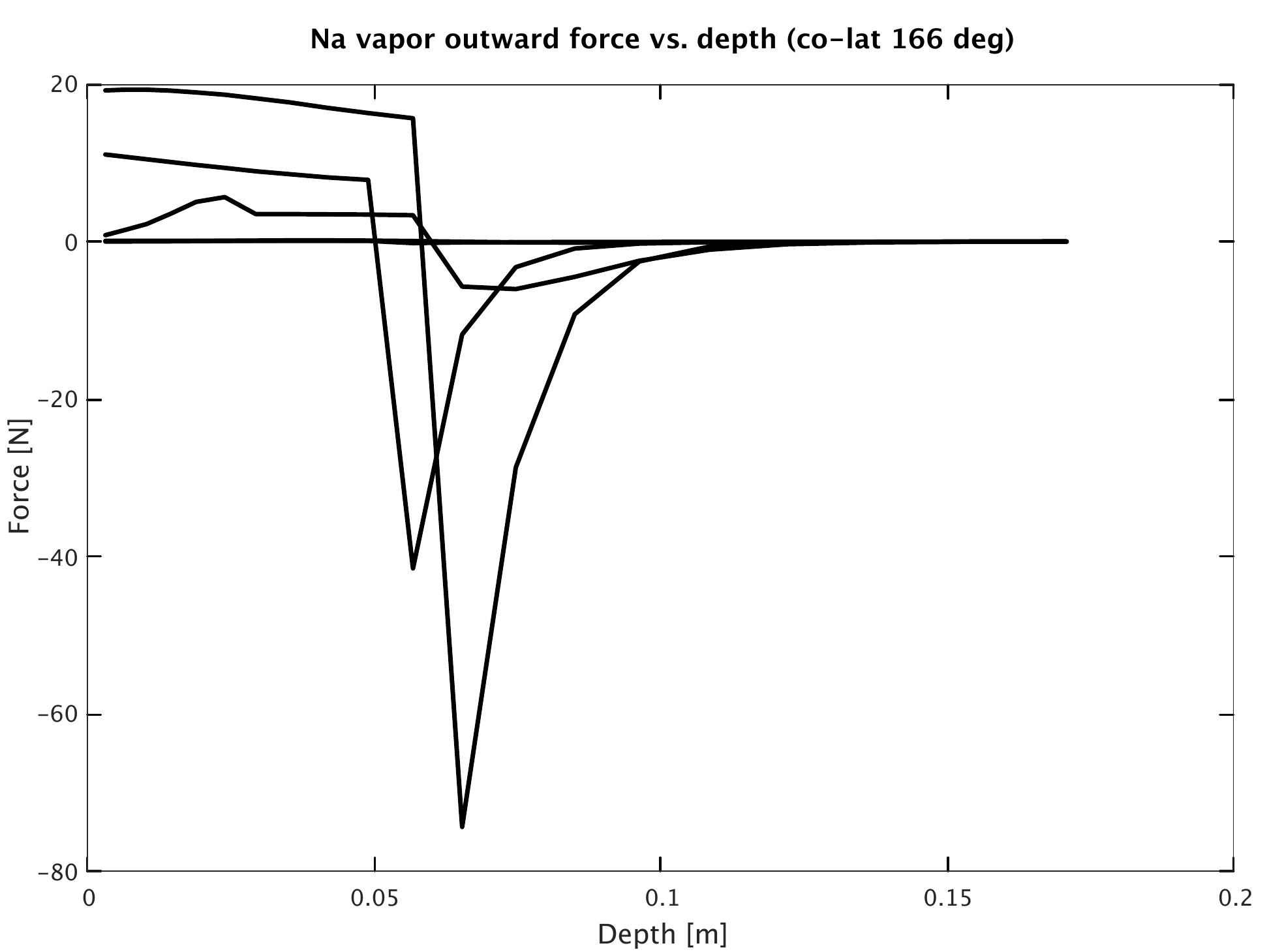}} & \\
\end{tabular}
     \caption{Upper left: The total sodium production rate calculated
       for Phaethon, as a function of heliocentric distance within
       $\sim 1\,\mathrm{AU}$ of the Sun (from 45 days
       pre--perihelion). Upper right: the location of the sodium
       sublimation front as a function of the angular distance from
       the north pole (the co--latitude) at perihelion. Middle left:
       the sodium abundance as a function of depth at latitude
       $76^{\circ}\,\mathrm{S}$ at perihelion, normalized to the
       initial abundance.  Middle right: the pressure created
         by sodium volatilization as a function of depth for a few
         rotational phases for $76^{\circ}\,\mathrm{S}$ at perihelion.
         Lower left: the corresponding force caused by the sodium
         pressure gradient shown in the middle right figure.
     }
     \label{fig.model2}
\end{figure}
\clearpage

Figure~\ref{fig.model2} shows that the temperatures experienced by
Phaethon are sufficiently high to cause significant sodium
volatilization, if the element is present in pure form.  Within $\sim
1\,\mathrm{AU}$ of the Sun, the sodium production rate increases six
orders of magnitude, briefly reaching a molar perihelion production
rate that parallels the water production rate of Comet
67P/Churyumov--Gerasimenko at $2.1\,\mathrm{AU}$ \citep{fougere16}. If
pure sodium were present and such strong outgassing takes place on
Phaethon, it would readily cause comet--like ejection of dust into
space. Figure~\ref{fig.model2} shows that the sodium sublimation front
withdraws rapidly below the surface, reaching a depth of
$0.4\,\mathrm{m}$ at the north pole at perihelion. The simulation did
not explicitly model the erosion of the surface because of comet--like
activity, therefore these depths should be considered upper limits on
the front withdrawal per orbit.

The pressure peak at the sub--surface
sublimation front forces sodium vapor to flow both outward and inward.
The outward flow is responsible for the outgassing into space, whereas
vapor flowing inwards eventually reaches colder material and
recondenses. This causes a gradual build--up of sodium at depth. The
\texttt{NIMBUS} simulations show that local concentrations can form
that are at least twice as high as the original abundance. Such
redistribution of solid sodium, especially when occurring slowly over
time and when combined with impacts that expose deeper layers to
space, could cause strong episodic bursts of activity.
Figure~\ref{fig.model2} also shows the force
$\mathbf{F}=-dp/dr\,\hat{r}$ caused by the pressure variation $p=p(r)$
with the radial coordinate $r$. The outward force here reaches a peak
of $\sim 20\,\mathrm{N}$. Given the low surface gravity of Phaethon
($g\approx 0.001\,\mathrm{m\,s^{-2}}$), such a force would be capable
of lifting meter--sized boulders.  Note that these results assume a
pure sodium phase just below the surface, and thus is meant to represent 
a bounding case.

Pure sodium is not common in carbonaceous chondrites and most of the
element is hosted by minerals like feldspar, nepheline
[(Na,K)AlSiO$_4$], sodalite and saponite \citep{rubin97}. However,
there are at least two mechanisms that could segregate sodium from the
host and create a pure phase in bodies like Phaethon that comes
unusually close to the Sun. Firstly, \citet{capek09} assembled atomic
diffusivity data for sodium for a range of relevant minerals and
demonstrated that substantial Fickian diffusion could occur at
temperatures reached by Phaethon on timescales similar to the orbital
period, if the grains are sufficiently small. Heating of material at
depths from which vapor transport to the surface is limited could
build a pure sodium deposit over time as the sodium gas cools
  and condenses to solid phase, that later could become exposed at
the surface by e.g., cratering or landslides. Secondly, laboratory
experiments by \citet{russell94} showed that layers of pure sodium
formed on the surface of $\mathrm{NaCl}$ particles when irradiated
with $0.5\,\mathrm{keV}$ electrons. Solar wind electrons typically
reach 10--$50\,\mathrm{eV}$, but superhalo electrons reach as much as
20--$200\,\mathrm{keV}$ \citep{wang10}.  The feasibility of sodium
separation from different types of host minerals due to space
weathering near the Sun should be studied further.  Our simulations of
sodium sublimation and outgassing from Phaethon should be considered a
``best--case scenario'' of how Phaethon's activity could behave in
optimum conditions. Under such circumstances Na--driven comet--like
activity seems feasible. However, continued research is needed to
better understand potential production mechanisms of pure sodium, the
strength of activity at lower abundances of sodium or when the sodium
content is present in different mineral phases, and the effect of
surface--layer processing on Phaethon during multiple orbits.

\section{Laboratory Analysis}

\subsection{Meteorite sample}

We obtained a sample of meteorite material to test its reaction to
high temperatures in order to determine the capability of sodium to be
a driver of activity on small Solar system bodies in a material
analogous to Phaethon.  For this study, we used cm-scale units of the
Allende (CV3) meteorite, totaling $67.94~$g in mass.  At the present,
there is no established analog for Phaethon or any Pallas-like B-type
asteroid in our meteorite collection.  We know that Phaethon
has a geometric visible albedo of $10-16\%$
\citep{hanus18,taylor19,masiero19}, which is lower than S-type
asteroids that have been linked to the ordinary chondrite (OC)
meteorites.  This link was established by the return of samples from
Itokawa by the Hayabusa mission \citep{hayabusa}, and thus we can
exclude OCs as a possible analog.  Phaethon is unlikely to be a metal
body and shows no similarity with Vesta or its family, which excludes
iron or HED meteorites as possible analogs.  As carbonaceous
chondrites are the largest group that cannot be excluded, we use them
for our study.

The choice of Allende specifically was driven by the need to acquire a
relatively large quantity of material away from the original surface
(and any potential terrestrial weathering) for our proposed
destructive analysis; as Allende was a large fall, it was the most
accessible carbonaceous chondrite for our study.  Allende is
classified as a CV3 chondrite and thus likely not a perfect match for
Phaethon, which has been suggested to be related to CM chondrites
\citep{hanus18epsc} or CK4 chondrites \citep{clark10},
although these associations are based on weak spectral
  features and so are uncertain.  Should Phaethon be a CK4 type, the
  metamorphic history experienced by CK4 chondrites would mean that
  plagioclase would be the most common sodium bearing mineral instead
  of sodalite and nepheline for Allende \citep{greenwood10}. However,
the goal of this work is to test the behavior of carbonaceous
chondrites to heating similar to what Phaethon experiences,
specifically their sodium-bearing minerals.  Future work, motivated by
the results of this study, will extend this investigation to other
carbonaceous meteorites that may be better analogs for Phaethon.

Previous work has shown that Allende contains an average of $0.46\%$
of Na$_2$O, making it the second most abundant volatile element after
sulfur at $\sim2.1~$wt$\%$ \citep{allende_ref}.  In contrast, evolved
comets like 67P/Churyumov-Gerasimenko have a $5-80\%$ mass fraction of
cometary volatiles in their nucleus \citep{fulle17,choukroun20}, in
this case water ice. This two-orders-of-magnitude difference would
imply that if sodium volatilization does spur activity on objects
close to the Sun, it would be expected to be at a much lower level
than the activity seen for comets, but potentially detectable with
sufficient sensitivity.  Water volatilization would be expected to be
significantly stronger, so sodium-driven activity would only be
possible for objects like Phaethon that do not contain subsurface
water.

To prepare the meteorite material for our tests, we crushed the sample
by hand in a mortar and pestle made of corundum, and filtered the
material through a $250~\mu$m sieve, reserving a few larger chips for
microscopic analysis.  We prepared 14 aliquants of powdered sample, each
$0.5~$g in mass.  We reserved the remaining material for future
analysis.  All tools and storage containers were cleaned and rinsed
with methanol prior to handling the sample.

\subsection{Experimental Setup}

The proposed temperature (1123 K) of Na release by sublimation of
sodalite by \citet{kasuga06} was derived from the $50\%$ condensation
temperature of Na in Solar nebula by
\citet{lodders03}. However, sodalite is a secondary mineral
  formed during metasomatic alteration of chondrites, meaning that the
  use of the condensation temperature of Na may not be fully
  applicable in this case.  Previous heating experiments of Allende
and Murchison for Na loss were conducted at 1323 - 1623 K
\citep{wulf95}. To evaluate if sublimation occurs at lower
temperatures as proposed by \citet{kasuga06} due to the
  prevailing conditions on asteroids near the Sun compared to at their
  formation, we tested a range of temperatures designed to span the
peak heating experienced by NEOs such as Phaethon.  We heated samples
of powder and chips at each of the following temperatures: $573~$K
($300~^\circ$C), $673~$K ($400~^\circ$C), $773~$K ($500~^\circ$C),
$873~$K ($600~^\circ$C), $973~$K ($700~^\circ$C), and $1073~$K
($800~^\circ$C).  We also held powder and chip samples in reserve as
controls.

Each aliquant was placed in a cleaned ceramic crucible, and then
placed inside an oven at room temperature.  The oven took
approximately 30-60 minutes to ramp up to the target temperature, and
then 30-120 minutes to cool down (with higher peak temperatures taking
longer to both heat and cool). The aliquant was held at peak
temperature for one hour. For simplicity, we refer to the heated
samples using the peak temperature.  The oven was at ambient
atmosphere during heating, which is much more oxidizing than the
heating in vacuum experienced by Phaethon. \citet{wulf95} and
\citet{sossi19} showed that Na volatility increases with decreasing
oxygen fugacity. Thus, any change we observed from our experiment
provides a lower bound on the Na loss in vacuum. The time-scale of
heating and cooling experienced by these samples roughly corresponds
to one diurnal cycle of Phaethon (which has a rotation period of
$3.6~$hr). Future work will investigate heating in a vacuum to better
simulate the environment Phaethon experiences.

After the experiment, one chip fragment from the
control sample (G) and each heated aliquant (A-F) were chosen randomly
to prepare a grain mount. These chips were embedded in
EpoCure$^{\rm{TM}}$ epoxy, and subsequently ground and polished to
generate a smooth flat surface for later analyses. Grinding
materials included SiC, diamond, and alumina. Except for diamond and
alumina in pore spaces and cracks, contamination by grinding materials
to the samples was not significant enough to be noticeable.

Chip A-G were imaged on a Hitachi SU3500 scanning electron
  microscope (SEM) equipped with an Oxford 150 energy dispersive
X-ray spectroscopy detector at the Jet Propulsion Laboratory
(JPL). Backscattered electron images (BSE) and maps were collected at
15 kV with a spot current intensity of 70. Each full area EDS map
mosaic is composed of a compilation of areas that were imaged and
analyzed at 150x magnification with image resolution of 1024 and 4
frames per map area. EDS full grain map mosaic images were converted
from text image files to png files in Fiji and images were colorized
and combined to generate overlay maps using Photoshop.

Precise compositions of different phases in the unheated Chip G and
heated Chip F were analyzed using a JEOL JXA-8200 electron probe
micro-analyzer (EPMA) at the California Institute of Technology. A
focused electron beam of 10 kV and 10 nA was used. The K$\alpha$ lines
of Si, Ti, Al, Cr, Fe, Mn, Mg, Ca, Na, K, and Cl were counted for 20 s
with Na being counted first to avoid significant Na loss during
analyses. The mean atomic number (MAN) method \citep{donovan96} was
used for the background corrections. Standards for analysis include
natural and synthetic materials.

\subsection{Experimental Results}

\begin{figure}[ht]
\centering
\includegraphics[scale=0.5]{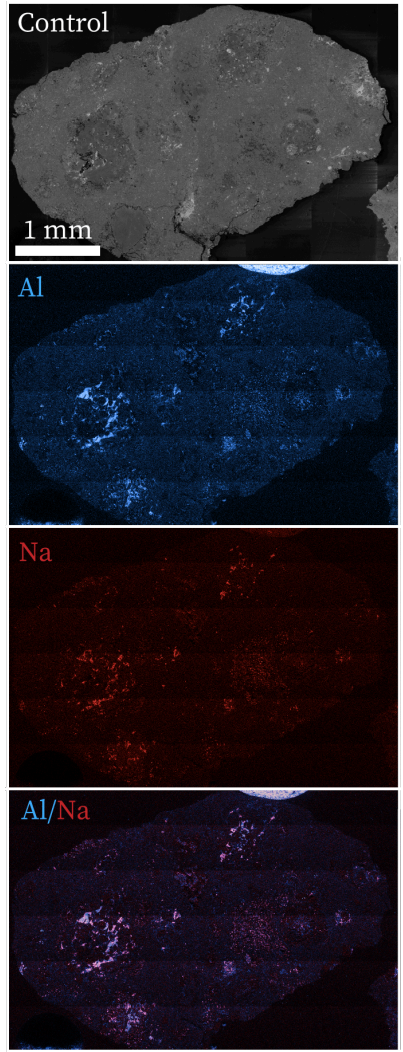}
\caption{Backscattered electron (BSE) image and the EDS Na and Al maps
  of the control sample (chip G). The Na and Al maps for the control
  and the heated chip (Fig \ref{fig.mapF}) were plotted on the same
  intensity scale to enable direct comparison.  An interactive figure
  is provided in the online version of this article, showing an
  overlay of the Na and Al maps, with a slidebar allowing the viewer
  to change between them.}
     \label{fig.mapG}
\end{figure}

\begin{figure}[ht]
\centering
\includegraphics[scale=0.5]{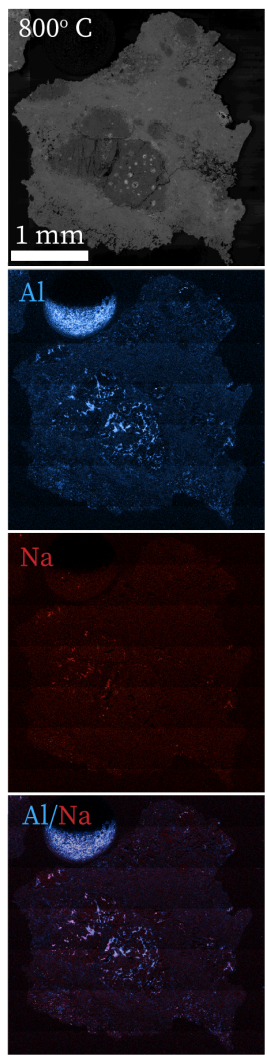}
\caption{The same as Fig~\ref{fig.mapG}, for the sample heated at
  $1073~$K (chip F). The bright, rounded feature near the top left of
  the EDS maps is an artifact from an epoxy bubble, and not part of the
  sample under investigation. An interactive figure is
    provided in the online version of this article, showing an overlay
    of the Na and Al maps, with a slidebar allowing the viewer to
    change between them.}
     \label{fig.mapF}
\end{figure}

The EDS full chip maps of aluminum and sodium for the unheated and
$1073~$K chips are presented in Figures \ref{fig.mapG} \&
\ref{fig.mapF}.  These maps show the distribution of elements present
in detectable abundances ($>0.1~$ wt$\%$). The EDS map of the unheated
chip G illustrates a close spatial association of Na with Al and Si
but not with Cl, consistent with an interpretation that the Na is
present in aluminosilicates, likely nepheline and feldspar.
Comparison of the EDS maps of the heated chips with the control sample
shows that only the sample heated to $1073~$K (chip F) displays lower
Na abundances than the control sample, whereas Al and Si are
comparable to the control sample. The areas around the phases that
should be sodium-rich in the heated sample do not show any increase in
sodium content in the EDS maps, suggesting that the reaction that
occurred was decomposition rather than Fickian diffusive loss.  This
suggests that Na-rich aluminosilicates in chip F lost Na when heated
to $1073~$K ($800~^\circ$C).

\clearpage

\begin{figure}[ht]
\begin{center}
\includegraphics[scale=0.7]{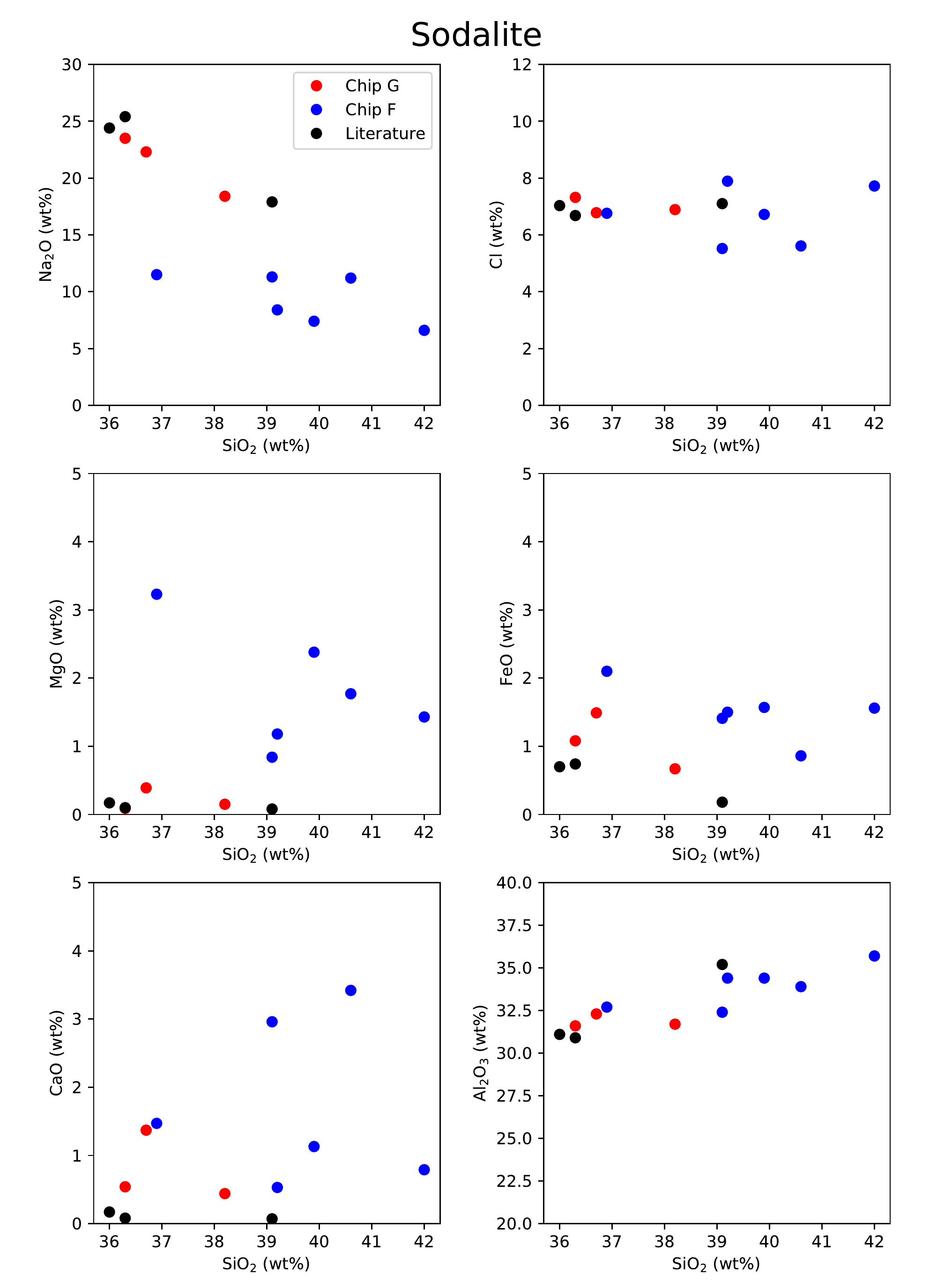} \protect\caption{EPMA data
  for sodalite phases comparing the fractional composition of various
  oxides to the SiO$_2$ weight percentage.  Shown are the control chip
  (Chip G), the heated chip (Chip F) and literature data from
  \citet{kimura95} and \citet{wasserburg11}.  Data show raw values
  that have not been renormalized to $100\%$. }
\label{fig.sodalite}
\end{center}
\end{figure}

\begin{figure}[ht]
\begin{center}
\includegraphics[scale=0.7]{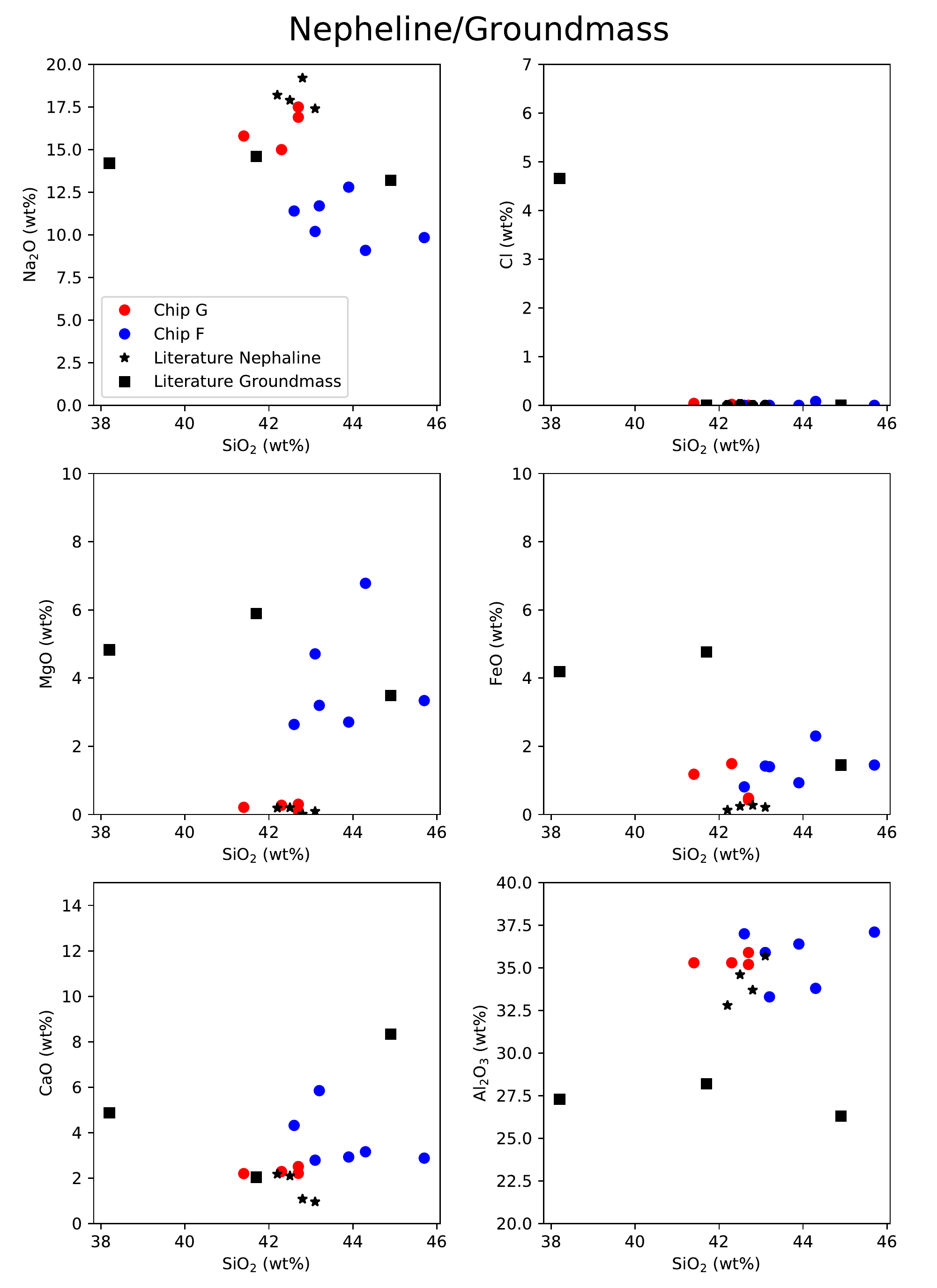} \protect\caption{The same as
  Figure~\ref{fig.sodalite}, but showing phases composed of nepheline
  or groundmass. }
\label{fig.neph_gm}
\end{center}
\end{figure}

EPMA data for Chip F and Chip G are listed in Table
\ref{tab.epma_soda} for sodalite phases and Table \ref{tab.epma_nef}
for nepheline and groundmass phases. Pyroxene and plagioclase about
$20~\mu$m away from the Na-rich phases did not show high Na content,
further supporting that the Na-rich phases did not lose Na through
Fickian diffusion. Note that after heating, sodalite and
  nepheline have been altered to different mineral phases. For
  simplicity, we refer to them as heated sodalite and nepheline.  Low
  oxide totals of heated sodalite in Chip F ($<97.6$ wt$\%$) could be
  caused by Na-loss, fine grains, or bad polish. Sodalite in unheated
  chip G is also fine grained and similarly polished, but oxide totals
  are acceptable. Therefore, we regard the bad totals of heated
  sodalite as mainly due to Na loss. Although these low totals are
  below expected values for complete analyses, we included them here
  to show the changes in Na contents. Compared to unheated sodalite in
  Chip G, heated sodalite Chip F clearly shows lower Na$_2$O and
higher Al$_2$O$_3$, CaO, and MgO, although Cl and FeO appear to be
similar between those two chips (Figure \ref{fig.sodalite}).
It is unclear why Cl did not show comparable changes, though
  it may be due to the low overall Cl content (see Supplementary Figs
  \ref{fig.Cl_G} and \ref{fig.Cl_F}). Further studies are needed to
  investigate the behavior of Cl in sodalite during heating.  The
heated nepheline in Chip F contains much higher MgO and CaO than
unheated ones in Chip G, although Al$_2$O$_3$ contents are comparable
between two chips (Figure \ref{fig.neph_gm}).  Increase in
  refractory components indicates that a small amount of mafic phases
  was annealed with the heated sodalite. However, the dilution effects
  by this mixing are inadequate to explain the decrease in Na.  The
heated nepheline in Chip F may be groundmass, given that its MgO and
CaO contents are similar to altered groundmass in \citet{kimura95}.
Regardless, Na$_2$O contents in heated nepheline in Chip F are much
lower than unheated nepheline in Chip G or unheated groundmass in
\citet{kimura95}. Using average values, heated sodalite and nepheline
in Chip F show maximum Na loss of $55\%$ and $35\%$,
respectively. These results demonstrate that at peak temperature
($1073~$K) for a relatively short time ($\sim1~$hour), Na can be lost
from nepheline and sodalite.  Although the changes in Chip F may be
due to thermal decomposition under our experimental conditions, these
results show that heating under vacuum may facilitate the generation
of Na vapor, which will be explicitly tested in future work.

\begin{table}[hb]
  \begin{center}
    \caption{Electron Probe Micro-Analyzer Results - Sodalite}
  \scriptsize{
\begin{tabular}{l|cccccc|ccc|ccc}
\hline
Oxide & \multicolumn{6}{c}{Chip F} & \multicolumn{3}{c}{Chip G} & \multicolumn{2}{c}{K95$^\dagger$} & W11$^\dagger$  \\
\hline
SiO$_2$ & 39.1 & 40.6 & 39.2 & 42.0 & 39.9 & 36.9 & 38.2 & 36.3 & 36.7 & 36.3 & 36.0 & 39.1\\
TiO$_2$& $<$0.03 & 0.04& $<$0.03& $<$0.03& $<$0.03 & 0.08& $<$0.03& $<$0.03& $<$0.03& $<$0.03& $<$0.03& $<$0.03\\
Al$_2$O$_3$ & 32.4 & 33.9 & 34.4 & 35.7 & 34.4 & 32.7 & 31.7 & 31.6 & 32.3 & 30.9 & 31.1 & 35.2\\
Cr$_2$O$_3$& $<$0.03 & 0.05 & 0.03& $<$0.03 & 0.05 & 0.04& $<$0.03& $<$0.03 & 0.06& $<$0.03& $<$0.03& $<$0.03\\
FeO & 1.41 & 0.86 & 1.50 & 1.56 & 1.57 & 2.10 & 0.67 & 1.08 & 1.49 & 0.74 & 0.70 & 0.18\\
MnO& $<$0.02& $<$0.02 & 0.09 & 0.14 & 0.04 & 0.22 & 0.03& $<$0.02& $<$0.02& $<$0.02& $<$0.02& $<$0.02\\
MgO & 0.84 & 1.77 & 1.18 & 1.43 & 2.38 & 3.23 & 0.15 & 0.09 & 0.39 & 0.10 & 0.17 & 0.08\\
CaO & 2.96 & 3.42 & 0.53 & 0.79 & 1.13 & 1.47 & 0.44 & 0.54 & 1.37 & 0.08 & 0.17 & 0.07\\
Na$_2$O & 11.3 & 11.2 & 8.4 & 6.6 & 7.4 & 11.5 & 18.4 & 23.5 & 22.3 & 25.4 & 24.4 & 17.9\\
K$_2$O & 0.10 & 0.07 & 0.03 & 0.05 & 0.06 & 0.14 & 0.12 & 0.03 & 0.04 & 0.04 & 0.02& $<$0.02\\
Cl & 5.52 & 5.61 & 7.89 & 7.72 & 6.72 & 6.76 & 6.89 & 7.32 & 6.78 & 6.68 & 7.03 & 7.10\\
Total & 93.6 & 97.6 & 93.4 & 96.0 & 93.6 & 95.1 & 96.6 & 100.5 & 101.4 & 100.2 & 99.6 & 99.6\\
\hline
\end{tabular}
\label{tab.epma_soda}
$^\dagger$Literature references: K95=\citet{kimura95}, W11=\citet{wasserburg11}
}
\end{center}
\end{table}

\begin{sidewaystable}
  \begin{center}
    \caption{Electron Probe Micro-Analyzer Results - Nepheline/Groundmass} 

  \scriptsize{
\begin{tabular}{l|cccccc|cccc|cccc|ccc}
\hline
Oxide & \multicolumn{6}{c}{Chip F} & \multicolumn{4}{c}{Chip G} & \multicolumn{3}{c}{K95$^\dagger$} & W11$^\dagger$ & \multicolumn{3}{c}{K95$^\dagger$} \\
\hline
 & \multicolumn{6}{c}{Nepheline/Groundmass} & \multicolumn{4}{c}{Nepheline} & \multicolumn{4}{c}{Nepheline} & \multicolumn{3}{c}{Groundmass} \\
\hline
SiO$_2$ & 42.6 & 43.9 & 43.2 & 45.7 & 43.1 & 44.3 & 42.7 & 42.7 & 42.3 & 41.4 & 42.5 & 42.2 & 42.8 & 43.1 & 44.9 & 41.7 & 38.2\\
TiO$_2$& $<$0.03& $<$0.03 & 0.05& $<$0.03& $<$0.03& $<$0.03 & 0.07& $<$0.03& $<$0.03& $<$0.03& $<$0.03& $<$0.03& $<$0.03& $<$0.03 & 0.27 & 0.32 & 0.36\\
Al$_2$O$_3$ & 37.0 & 36.4 & 33.3 & 37.1 & 35.9 & 33.8 & 35.9 & 35.2 & 35.3 & 35.3 & 34.6 & 32.8 & 33.7 & 35.7 & 26.3 & 28.2 & 27.3\\
Cr$_2$O$_3$& $<$0.03 & 0.04& $<$0.03& $<$0.03& $<$0.03& $<$0.03 & 0.06 & 0.09& $<$0.03& $<$0.03& $<$0.03 & 0.16& $<$0.03& $<$0.03 & 0.65 & 0.09 & 0.25\\
FeO & 0.81 & 0.93 & 1.40 & 1.45 & 1.42 & 2.30 & 0.43 & 0.48 & 1.49 & 1.18 & 0.24 & 0.13 & 0.27 & 0.21 & 1.45 & 4.77 & 4.19\\
MnO & 0.11 & 0.06 & 0.03 & 0.39& $<$0.02& $<$0.02 & 0.02& $<$0.02& $<$0.02& $<$0.02 & 0.07 & 0.06& $<$0.02 & 0.04 & 0.03 & 0.00 & 0.14\\
MgO & 2.64 & 2.71 & 3.20 & 3.34 & 4.71 & 6.78 & 0.10 & 0.30 & 0.27 & 0.21 & 0.20 & 0.19 & 0.01 & 0.09 & 3.49 & 5.89 & 4.83\\
CaO & 4.32 & 2.93 & 5.85 & 2.88 & 2.79 & 3.16 & 2.21 & 2.51 & 2.29 & 2.20 & 2.10 & 2.18 & 1.08 & 0.96 & 8.33 & 2.04 & 4.88\\
Na$_2$O & 11.4 & 12.8 & 11.7 & 9.84 & 10.2 & 9.09 & 16.9 & 17.5 & 15.0 & 15.8 & 17.9 & 18.2 & 19.2 & 17.4 & 13.2 & 14.6 & 14.2\\
K$_2$O & 0.44 & 0.49 & 0.37 & 0.29 & 0.24 & 0.17 & 2.18 & 2.12 & 1.92 & 1.91 & 2.09 & 1.82 & 1.93 & 1.50 & 1.27 & 1.56 & 0.01\\
Cl& $<$0.02& $<$0.02& $<$0.02& $<$0.02& $<$0.02 & 0.08& $<$0.02& $<$0.02 & 0.02 & 0.04 & 0.02& $<$0.02& $<$0.02& $<$0.02& $<$0.02& $<$0.02 & 4.66\\
Total & 99.4 & 100.2 & 99.1 & 101.0 & 98.3 & 99.6 & 100.5 & 101.0 & 98.6 & 98.1 & 99.7 & 97.7 & 98.9 & 99.0 & 99.8 & 99.2 & 99.0\\
\hline  
\end{tabular}
\label{tab.epma_nef}
$^\dagger$Literature references: K95=\citet{kimura95}, W11=\citet{wasserburg11}
}
\end{center}
\end{sidewaystable}

\clearpage

\section{Discussion}

Nepheline in carbonaceous chondrites is a secondary mineral formed
through thermal metamorphism or metasomatism of primary materials
\citep[e.g.][]{ikeda95,kimura95,krot98}. Another mineral,
sodalite [Na$_4$(Si$_3$Al$_3$)O$_{12}$Cl], could also be a source for
Na volatilization.  Nepheline and sodalite in carbonaceous chondrites
typically occur as accessory phases in trace quantities, but samples
with high abundances are present \citep[e.g.][]{wasserburg11}. Surface
materials on Phaethon could possibly contain significant abundances of
nepheline or sodalite that would contribute to the observed
vaporization event.

Previous heating experiments of Allende conducted by \citet{wulf95}
reached higher temperatures than those in our study, with samples
heated for longer times.  As such, those experiments probed a
different regime of thermal exposure than our work, more
  appropriate to the study of silicate vaporization. In those
experiments, powder and chips of Allende and Murchison were heated to
$1323-1623~$K ($1050-1350~^\circ$C) under different oxygen
fugacities. Results by \citet{wulf95} showed that more Na is lost
under reduced conditions. Na vaporization can be written in a
generalized reaction form: $\frac{1}{2}$ Na$_2$O (s or l) = Na (g) +
$\frac{1}{4}$ O$_2$ (g). Thus, the lower oxygen fugacity drives this
reaction toward right hand side.  At $1323~$K, only $26\%$ bulk Na was
lost from samples heated in air. In contrast, our experiments showed
up to $55\%$ loss of Na from sodalite and nepheline. We attribute this
difference to the fact that Na is hosted in this sample in both
refractory grains (albite) and secondary phases (nepheline and
sodalite).  While our analysis focused only on the secondary phase
minerals, the bulk analysis contains both refractory and secondary
phases and therefore would have a more muted sodium loss.  More
importantly, our experimental results demonstrate that Na can be lost
from secondary minerals such as nepheline in a short heating period
that is predicted from the thermophysical modeling above. Thus the
observed losses combined with the predictions from models support our
theory that sodium can be a driver of activity on small, near-Sun
asteroids.

The OSIRIS-REx spacecraft has observed recently-launched
particles in short-term stable orbits around the carbonaceous
NEO (101955) Bennu \citep{lauretta19}.  These
centimeter-scale objects were observed to be ejected from the surface,
typically in the late afternoon on low-energy orbits.  Numerous
possible ejection mechanisms including thermal fracturing,
electrostatics, and phyllosilicate dehydration were investigated, but
no direct cause was confirmed.  The surface of Bennu does not reach
the same peak temperatures as that of Phaethon, making sodium
volatilization unlikely to be responsible for these events.  Still,
Bennu stands as another interesting example of unexpected forms of
activity on NEOs in addition to Phaethon.

Recent work by \citet{maclennan20} describe the long-term evolution of
the thermal history of Phaethon, showing that it experiences cyclical
changes to its maximum heating from the Sun.  Based on those results,
Phaethon currently is just past the peak of its $\sim18~$kyr heating
cycle, which would imply that sodium volatilization should be ramping
down.  This heating history is consistent with Phaethon having had
higher levels of sodium-driven activity in the past few thousand
years. This could then explain how Phaethon created the Geminid meteor
stream in the recent past, but has a present day activity insufficient
to produce it.  Additional models by \citet{maclennan20} predict that
asteroid (155140) 2005 UD, which may have been formed from an earlier
splitting of Phaethon, is at the beginning of a new heating cycle, and
we would expect sodium volatilization to increase over the next
$\sim5~$kyr from a currently quiescent state.

When considering the broader NEO population, \citet{granvik16}
demonstrated that there is a significant lack of low albedo asteroids
with very low perihelion distances.  They attribute this lack of
small, dark objects to extreme thermal events, though for many of the
asteroids with low perihelia the time to reach these orbits is
non-trivial.  We would therefore not expect a breakup scenario like
what is observed for low-perihelia comets, as these NEOs spend a
significant time heated to well above the temperature for ice
sublimation.  Any existing ice must be deeply buried and would not
respond rapidly (or catastrophically) to heating.  Sodium
volatilization, then, could act as the needed trigger to remove some
of the dehydrated surface material and expose enough ice-rich
subsurface to trigger the catastrophic breakups needed to explain the
observations of \citet{granvik16}.

\section{Conclusions}

Both our theoretical thermophysical modeling as well as our laboratory
analyses indicate that sodium has the potential to mobilize dust on
the surface of small bodies when other more volatile agents
(e.g. water) are not present.  This mechanism could then explain the
activity seen on objects with very low perihelia such as Phaethon.
However, this activity would depend strongly on the mineral phases
present and the specific surface geology of these objects.

Further work is needed to make more specific predictions for Phaethon,
including conducting heating experiments in a vacuum and using
meteorite materials that are closer analogs to Phaethon than Allende.
Likewise, future investigations of larger quantities of heated analog
material would enable testing of the efficiency of deposition of a
pure sodium layer that could later be rapidly triggered to cause
outgassing.  Results of these studies then would lead to improved
models of dust mobilization on the Phaethon's surface.

The DESTINY+ mission that is currently being pursued by JAXA
\citep{arai18,arai19} will provide {\it in situ} observations of
Phaethon for the first time.  These data may allow us to determine if,
and to what extent, sodium contributes to the observed activity on
Phaethon.  If Phaethon's dust emission is indeed driven by sodium
desorption, this mechanism provides another avenue for mass loss on
small Solar system bodies, especially those close to the Sun.

More broadly, simulations of the orbital evolution of the whole
NEO population by \citet{marchi09} indicate that
approximately two percent of the current NEO population
had or will have a perihelion distance within 0.1 AU at some point
their dynamical lifetimes.  This means that sodium volatilization
would not be expected to be restricted to just Phaethon, but instead
may drive activity on a small but important subset of the whole
NEO population.  Investigations of other asteroids with
small perihelia will allow us to determine the extent to which this
novel form of activity is present among the asteroids.

\section*{Acknowledgments}

The authors would like to thank Vishnu Reddy for assistance in
acquiring the meteorite samples used for this work. We thank the
anonymous referees for comments that helped improve this work.  This
research was carried out at the Jet Propulsion Laboratory, California
Institute of Technology, under a contract with the National
Aeronautics and Space Administration (80NM0018D004).  Funding for this
work was provided by a JPL Lew Allen award.

\appendix

\renewcommand\thefigure{\thesection.\arabic{figure}}    
\setcounter{figure}{0}

\section{Supplemental Figures}

\begin{figure}[ht]
\centering
\includegraphics[scale=0.35]{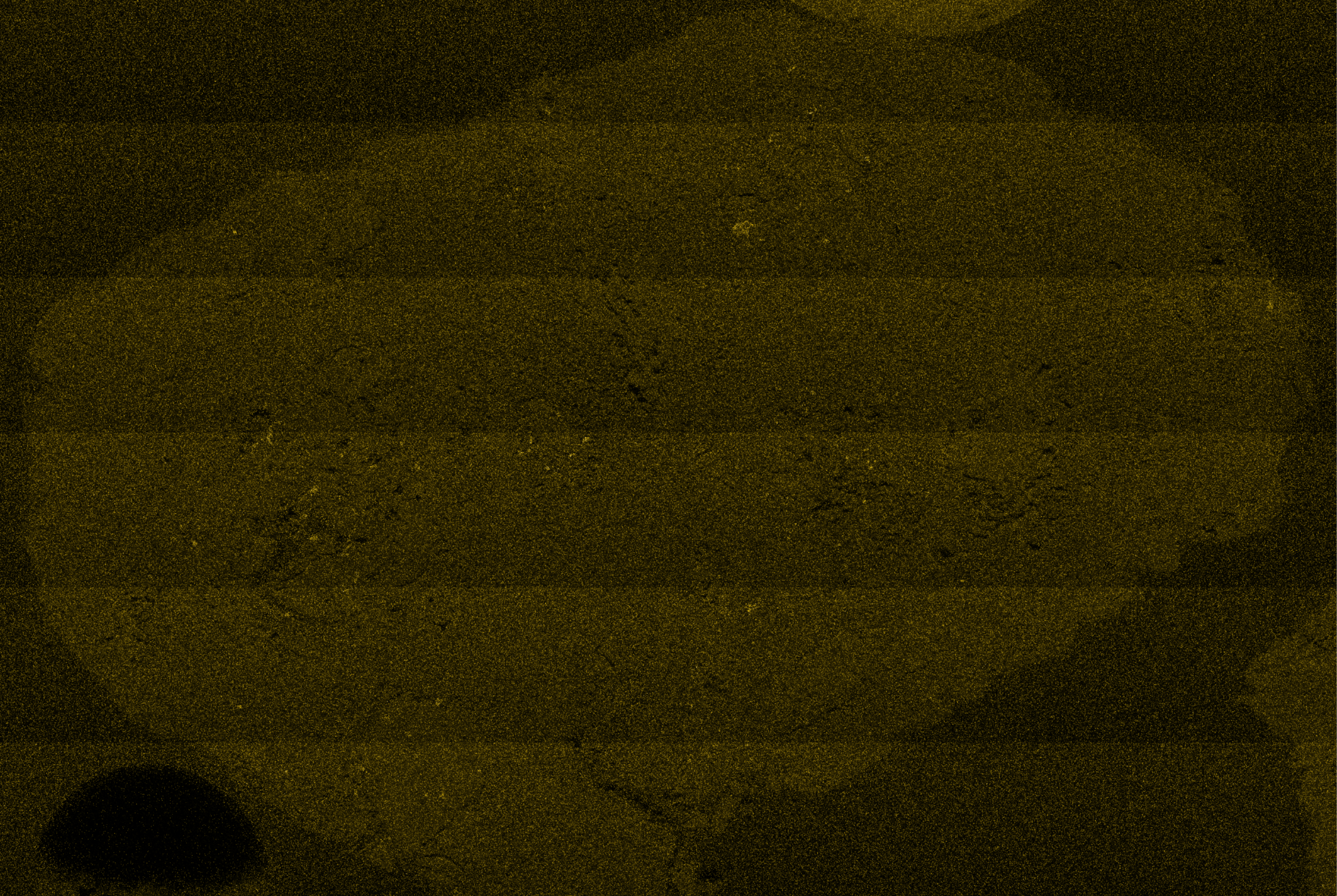}
\caption{EDS Cl map of the control sample (chip G), for comparison with the Na and Al maps shown in Fig~\ref{fig.mapG}. }
     \label{fig.Cl_G}
\end{figure}

\begin{figure}[ht]
\centering
\includegraphics[scale=0.35]{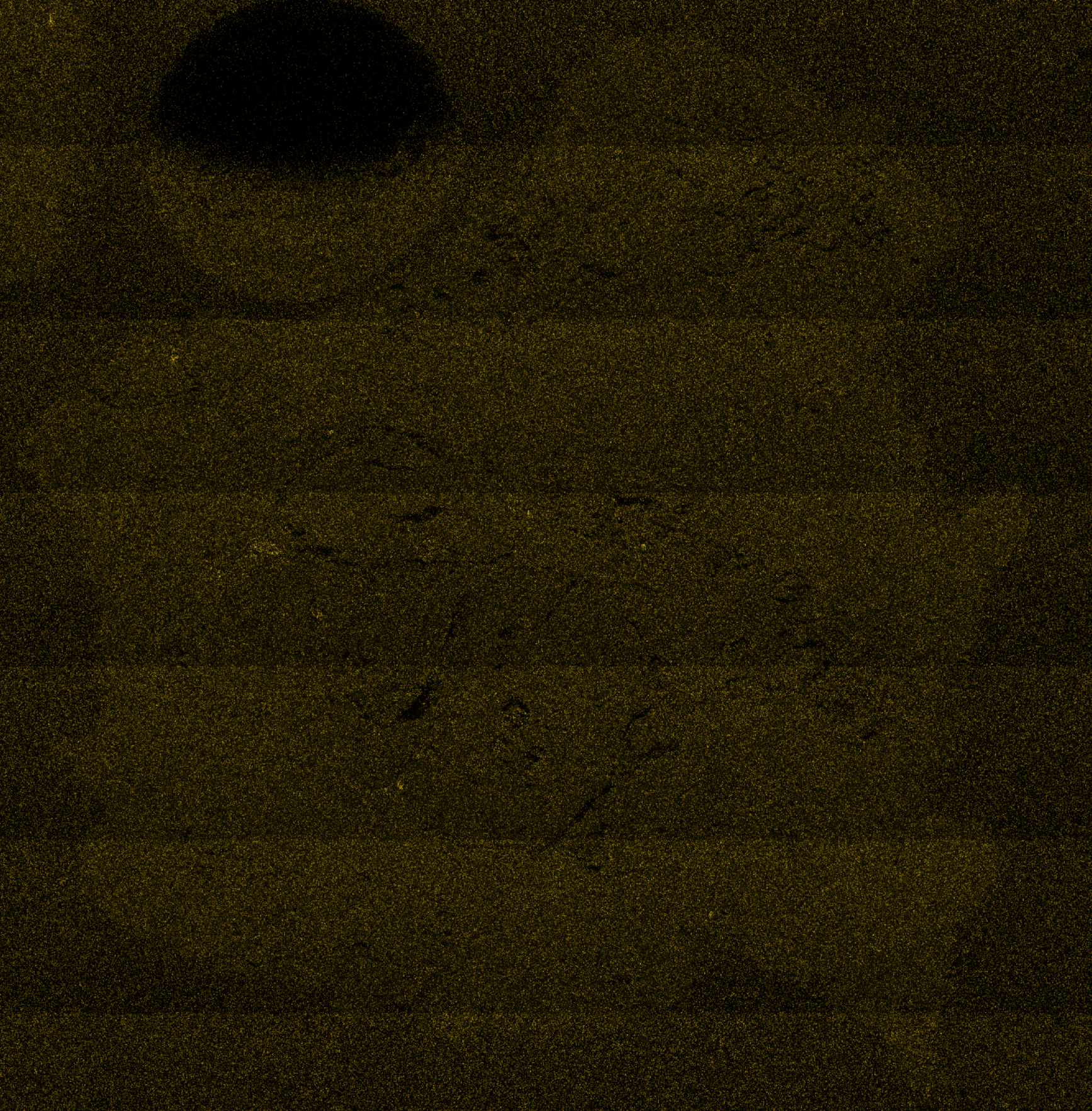}
\caption{EDS Cl map of the sample heated at $1073~$K (chip F), for
  comparison with the Na and Al maps shown in Fig~\ref{fig.mapF}. }
     \label{fig.Cl_F}
\end{figure}
\clearpage

\end{document}